\documentclass[12pt]{article}
\usepackage{epsf}
\usepackage{amsmath}

\usepackage{graphics}
\usepackage{cite}

\begin{titlepage}

\begin{document}

\hfill January 2020

\begin{center}

{\bf \Large Particle Theory at Chicago in the Late Sixties and p-Adic Strings\\
}
\vspace{2.5cm}
{\bf Paul H. Frampton\\  }
\vspace{0.5cm}
{\it Dipartimento di Matematica e Fisica "Ennio De Giorgi",\\ 
Universit\`{a} del Salento and INFN-Lecce,\\ Via Arnesano, 73100 Lecce, Italy\footnote{paul.h.frampton@gmail.com}
}

\vspace{1.0in}

\begin{abstract}
\noindent
As a contribution requested by the editors of a Memorial Volume for
Peter G.O. Freund (1936-2018) we recall the lively particle theory group
at the Enrico Fermi Institute of the University of Chicago in the late
sixties, of which Peter was a memorable member. We also discuss
a period some twenty years later when our and Peter's research overlapped
on the topic of p-Adic strings.

\end{abstract}

\bigskip
\bigskip
\bigskip
\bigskip
\bigskip

{\bf In memory of Peter G.O. Freund}

\end{center}

\vspace{1.2in}

\noindent
Submitted to a special issue of Journal of Physics A entitled \\
{\it A passion for theoretical physics: in memory of Peter G.O. Freund}. \\Edited by
J. Harvey, E. Martinec and R. Nepomechie.

\end{titlepage}

\section{Introduction} 

\noindent
It is a pleasure and an honour to write in memory of
Professor Peter G.O. Freund (1936-2018) who spent most of his career,
from 1965 to 2018, at the University of Chicago. As can be read elsewhere
in this book he had been born and initially educated in Romania
then acquired his PhD at the University of Vienna which explains his
European sophistication. He had a passion
for theoretical physics and inspired many young people, including
myself, in their quest to extend human knowledge.

\bigskip

\section{Enrico Fermi Institute}

\bigskip

\noindent
At the beginning of September 1968, I arrived at the University of
Chicago group in the Enrico Fermi Institute for a two-year postdoctoral position
having just finished a D.Phil in Oxford. I had a happy and productive time
within the following group of particle theorists:
\noindent
{\it Faculty:} Peter Freund, Yoichiro Nambu, Reinhard Oehme, J.J. Sakurai.
\noindent
{\it Postdocs:} Lay Nam Chang, Paul Frampton, Bodo Hamprecht, Ray Rivers, Joe Scanio. 

\bigskip

\noindent
Every weekday, most of the group would attend
a meeting for tea where new developments in particle theory were
discussed. On one day a week there was typically a seminar by an
external speaker. On Thursday afternoons, there was mandatory
attendance (mentioned in the offer letter) at an Enrico Fermi seminar where
a speaker was chosen from the attendees without prior warning. 

\bigskip

\noindent
After I arrived we discussed the recently-discovered Veneziano
model, especially with Nambu at first then the year later with Freund.
The progress by Nambu in string theory was remarkable. By the end of 
1968, after some early help by me with the 4-point function
he had factorised the N-point function and derived 
its exponential degeneracy. The Veneziano
model was discovered too late for my 1968 Oxford thesis on finite-energy
sum rules but was used in a similar follow-up Oxford D.Phil in 1969 by Michael
Kosterlitz with the same supervisor, J.C. Taylor. Kosterlitz later went
on to be awarded a Nobel prize in 2016 for very different research
in condensed matter theory.

\bigskip

\noindent
Before discussing my joint work with Peter, just to memorialise
the daily tea meetings, let me recall the four faculty, all by now
unfortunately deceased. Oehme (1928-2010) was a formal
expert on quantum field theory and did not
usually participate in socialising but was accessible in his
office for intense discussions. Sakurai (1928-1982) was
a true phenomenologist who favoured the idea that photon
interactions with hadrons are dominated by vector mesons,
including in a paper we published jointly, later at UCLA
\cite{PHFJJS}. Nambu (1921-2015) had a legendary reputation
from several major accomplishments,
including the discovery of spontaneous symmetry breaking
in field theory for which he won a Nobel Prize in 2008. I discussed
the Veneziano model with him\cite{NambuFestschrift} and
we published one paper together\cite{PHFYN}.

\bigskip

\noindent
Peter Freund was the youngest and most gregarious
of the faculty in the group. Every afternoon he had some
new idea to discuss with anybody willing to listen. His
enthusiasm was contagious and he produced many
publications. It was said he could conceive of an idea,
do the relevant calculations and write up a paper all
in one day.
Certainly he had the energy of a dynamo but the flexibility
to drop any idea which was demonstrably false. I did not
collaborate with him during my first year because I was
discussing the bosonic string with Nambu almost every
weekday, as well as writing other papers with co-postdocs.

\bigskip

\noindent
It seemed inevitable that in my second year we did coauthor
two interesting papers. The Veneziano model could be interpreted as a  
possible description of the duality in scattering of two mesons
each composed of a $(q \bar{q})$ pair of quark-antiquark.
The question addressed in \cite{PHFPGOF,EFFG} was how to
extend the description to baryons ($qqq$) and exotic hadrons
such as $qq\bar{q}\bar{q}$, $qqqq\bar{q}$, etc. It was argued
that the additional quarks are focused to a point on the string
worldsheet. Explicit formulas were suggested to describe these
situations. In \cite{EFFG}, an imaginative construction based on
a {\it Star-of-David} rule was hypothesised.

\bigskip

\noindent
These two papers were written at an early time (1970) when the
string picture was not fully developed but I do remember how much
fun we had in writing and publishing them and they well illustrated
the creative imagination of Freund who generated ideas
while the rest of us did calculations.

\bigskip

\noindent
According to my list, I coauthored work either at Chicago or later
with all the other four postdocs in the group
\cite{Chang,Hamprecht1,Hamprecht2,Rivers,Scanio} which
illustrates how Peter Freund successfully encouraged such
meetings of the minds.

\bigskip

\section{p-Adic strings}

\bigskip

\noindent
Starting from 1968, string theory experienced an up-and-down
history. After the enthusiasm and support for dual models of
strong interactions from 1968 to 1974 interest fell off in favour
of QCD as the correct strong
interaction theory.
But as an attempt to unify all elementary particle
forces including gravity, superstrings made a revival during 1984-87
when anomaly-cancellation and Calabi-Yau spaces appeared
to make a major breakthrough feasible. By 1987, such a dramatic
contact to the real world, for example derivation of one or more
of the many parameters
in the standard model, became less likely
because string theory, especially its numerous possible
compactifications, itself was shown to contain as many or more parameters.

\bigskip

\noindent
In 1987, by coincidence, Peter and I simultaneously noticed a surprising paper
\cite{Volovich} coming from Russia which made a curious new
observation about the Veneziano model. By expressing the
Euler gamma functions therein in terms of Riemann zeta
functions, different p-adic strings appeared, one for each
prime number $p$. This impressed also Witten and my postdoc
Okada and the Chicago and UNC groups began a friendly
competition to understand p-adic strings better. One motivation
was surely the appearance of algebraic number theory and
the excitement generated by the idea that prime numbers
could be useful in theoretical physics.

\bigskip

\noindent
Freund and Witten \cite{FWitten}
showed that the adelic infinite product
over the p-adic 4-point functions gave the original Veneziano
formula.

\bigskip

\noindent
With the superscript $(p)$ denoting a prime number and
the convention that $p=\infty$ denotes the real number field
so that $Q_p \rightarrow {\cal R}$ as $p \rightarrow \infty$
the Veneziano model is

\begin{equation}
A_0^{(\infty)} (s, t, u) 
= \int_0^1 dx |x|^{-\alpha(s)-1} |1 - x|^{-\alpha(t)-1}
= \frac{ \Gamma(-\alpha(s)) \Gamma(-\alpha(t))}{\Gamma(-\alpha(s)-\alpha(t))}
\label{Veneziano}
\end{equation}
where $\alpha(s)= 1 + \frac{1}{2} s$, $s+t+u=-8$ and therefore
$\alpha(s)+\alpha(t)+\alpha(u) = -1$.

\bigskip

\noindent
Adding the terms related by crossing symmetry
\begin{equation}
A^{(\infty)} =  A_0^{(\infty)}(s, t, u) + A_0^{(\infty)}(t, u, s) + A_0^{(\infty)}(u, s, t)
\end{equation}
and define $B^{(\infty)}$ by
\begin{equation}
A^{(\infty)}(s, t, u) = g_{\infty}^2 B^{(\infty)} (-\alpha(s), -\alpha(t))
\end{equation}
with
\begin{equation}
B^{(\infty)} (-\alpha(s),-\alpha(t)) = \int_{-\infty}^{+\infty} dx |x|^{\alpha(s)-1} |1-x|^{-\alpha(t)-1}
\end{equation}
in which the integration range is the real field ${\cal R} = Q_{\infty}$ rather than just $(0, 1)$.

\bigskip

\noindent
The p-adic string amplitude is defined by replacing ${\cal R}$ with the p-adic number field
$Q_p$.

\begin{eqnarray}
B^{(p)} (- \alpha(s), - \alpha(t)) & = & \int_{Q_p} dx |x|_p^{-\alpha(s)-1} |1-x|_p^{\alpha(t)-1}
\nonumber \\
&=& \Pi_{x=s,t,u} \left ( \frac{1 - p^{-\alpha(x)-1}}{1 - p^{+\alpha(x)}} \right) 
\label{padicstring}
\end{eqnarray}

\bigskip

\noindent
Recall the definition of the Riemann zeta function:

\begin{equation}
\zeta(z) = \Pi_p \left( \frac{1}{1-p^{-z}} \right) \equiv \Sigma_{r=1}^{\infty} \left( \frac{1}{r^z} \right)
\label{zeta}
\end{equation}
in order to rewrite
\begin{equation}
\Pi_p B^{(p)} ( - \alpha(s), - \alpha(t) ) = \Pi_{x=s,t,u} \frac{\zeta(-\alpha(x))}{\zeta(1 + \alpha(x))}
\end{equation}

\noindent
Using the relationship 
\begin{equation}
\Gamma(z)\zeta(z) = (2\pi)^{s-1}\sin (\frac{\pi z}{2}) \Gamma(1-z)\zeta(1-z)
\end{equation}
leads to the Freund-Witten adelic infinite-product formula
\begin{equation}
\Pi_p B^{(p)} ( - \alpha(s), - \alpha(t) ) = \left[ B^{(\infty)} ( - \alpha(s), - \alpha(t) ) \right]^{-1}
\label{FW}
\end{equation}

\bigskip

\noindent
Eq.(\ref{FW}) was their main result. The right-hand-side is (the inverse of) the Veneziano
model which is now seen to be equal to an infinite product of p-adic amplitudes.
This provided the initial excitement which led to a flurry of activity in 1988
because it suggested that the p-adic string in Eq.(\ref{padicstring}) might be
more basic than the bosonic string.

\bigskip

\noindent
The next step was to generalise to the
N-point functions\cite{FO1} for $N\geq5$, 

\bigskip

\noindent
Consider therefore the p-adic 5-point function

\begin{equation}
A_5^{(p)} - \int_{Q_p} dxdy |x|_p^{-\alpha_{35}-1} |1-x|_p^{-\alpha_{23}-1}
|y|_p^{-\alpha_{45}-1}|1-y|_p^{-\alpha_{24}-1}|x-y|_p^{-\alpha_{34}-1}
\end{equation}
where $\alpha_{ij} = 1+\frac{1}{2}(k_i+k_j)^2$.

\bigskip

\noindent
This double integral gives the additive form

\begin{eqnarray}
A_5^{(p)}& = & \Sigma_{(ij)(kl)} \left( \frac{ (1-p^{-1}) (1-p^{-1}) } {(1-p^{\alpha_{ij}})(1-p^{\alpha_{kl}}) } \right). \nonumber \\
& & -(2-p^{-1}) \Sigma_{(ij)} \left( \frac{(1-p^{-1})}{(1-p^{\alpha_{ij}}) } \right)
 + (2-p^{-1})(3-p^{-1})
\label{A5}
\end{eqnarray}
where the summations are over compatible (by duality) poles with $15$ and $10$ terms
respectively.

\bigskip

\noindent
For N=6, we find

\begin{eqnarray}
A_6^{(p)} & = & \Sigma \Pi_{i=1}^{i=3} \frac{ (1-p^{-1})}{(1-p^{\alpha_i})}
- (2-p^{-1}) \Sigma \Pi_{i=1}^{i=2} \frac{ (1-p^{-1})}{(1-p^{\alpha_i})} \nonumber \\
& & +(2-p^{-1})(3-p^{-1}) \Sigma_{ij} \frac{(1-p^{-1})}{(1-p^{\alpha_{ij}})} \nonumber \\
& & + (2-p^{-1})^2 \Sigma_{ijk} \frac{(1-p^{-1})}{(1-p^{\alpha_{ijk}})} \nonumber \\
& & - (2-p^{-1})(3-p^{-1})(4-p^{-1})
\label{A6}
\end{eqnarray}
where $\alpha_{ijk} = 1 + (k_i+k_j+k_k)^2$. The 1st and 2nd sums both have
105 terms; the 3rd and 4th have 15 and 10 terms respectively.

\bigskip

\noindent
When we examine factorisation of $N=5$, we find

\begin{eqnarray}
A_5^p  & \stackrel {\alpha_{12} \rightarrow 0}{\longrightarrow} &
\left( \frac{ (1-p^{-1})}{(1-p^{\alpha_{12}})}  \right) A_4^p.  \nonumber \\
& \sim &
 - \left( \frac{1 - p^{-1}}{\ln p } \right) \left( \frac{1}{\alpha_{12}} \right) A_4^p
\label{5fact}
\end{eqnarray}

\bigskip

\noindent
while, for $N=6$, at a 2-particle pole

\begin{eqnarray}
A_6^p  & \stackrel {\alpha_{12} \rightarrow 0}{\longrightarrow} &
\left( \frac{ (1-p^{-1})}{(1-p^{\alpha_{12}})}  \right) A_5^p.  \nonumber \\
& \sim &
 - \left( \frac{1 - p^{-1}}{\ln p } \right) \left( \frac{1}{\alpha_{12}} \right) A_5^p
\label{6fact}
\end{eqnarray}

\bigskip

\noindent
and at a 3-particle pole

\begin{eqnarray}
A_6^p  & \stackrel {\alpha_{123} \rightarrow 0}{\longrightarrow} &
A_4^p \left( \frac{ (1-p^{-1})}{(1-p^{\alpha_{123}})}  \right) A_4^p.  \nonumber \\
& \sim &
A_4^p  \left[ - \left( \frac{1 - p^{-1}}{\ln p } \right) \right] \left( \frac{1}{\alpha_{123}} \right) A_4^p
\label{6fact2}
\end{eqnarray}

\bigskip

\noindent
From formulas (\ref{5fact}), (\ref{6fact}) and (\ref{6fact2}), the prescription
for the p-adic propagator is

\begin{equation}
\frac{(1-p^{-1})}{(1-p^{\alpha(k^2)})}
\label{propagator}
\end{equation}

\bigskip

\noindent
and the p-adic degree $m$ vertex $V_m$ is

\begin{equation}
V_m= (-1)^{m+1} \Pi_{n=2}^{m-2} (n - p^{-1})
\label{vertex}
\end{equation}
for $m \geq 4$ and $V_3 = 0$.

\bigskip

\noindent
In \cite{FO1}, we were able to derive these Feynman rules
more quickly than the Chicago group\cite{BFOW} by starting
from the N-point function of \cite{BardakciRuegg} which is an
Eulerian integral of the first kind, as is Eq.(\ref{Veneziano}),
while the Chicago group began with the N-point function
of \cite{KobaNielsen} which is less explicitly so.

\bigskip

\noindent
For fixed $p$, the p-adic string can be described by a non-local scalar
field theory \cite{FO2} with lagrangian

\begin{eqnarray}
L_0 & = & \frac{1}{2} \Phi 
\left( \frac{1+p^{1+\partial\partial/2}}{1-p^{-1}} \right) \Phi
\nonumber \\
& & - \frac{1}{(p^{-1}+1)p^{-1}(p^{-1}-1)}
\left[ (1+\Phi)^{1+p^{-1}} -1 - (1+p^{-1}) \Phi \right. \nonumber \\
& & \left. - \frac{ ( 1+p^{-1})p^{-1}}{2} \Phi^2 \right]
\label{nonlocal}
\end{eqnarray}
which gives the Feynman rules discussed above.

\bigskip

\noindent
After 1988, because no new insight was being immediately gained the interest
in p-adic strings temporarily subsided, although it has been more recently
used in studying
the vacuum of the bosonic string\cite{Sen} and a modern
viewpoint is presented \cite{Gubser}
by the late Steven Gubser and coauthors in this Festschrift.

\section{Discussion}

\bigskip

\noindent
Peter Freund was a special personality in the EFI-Chicago group.
There were one German, Oehme, and two Japanese, Nambu
and Sakurai, all intellectually humble. As an intellectual,
Freund was unusually extrovert and animated
and always had new ideas which he was eager to discuss.

\bigskip

\noindent
Sakurai and Nambu were tied by their common language
and I recall one Japanese visitor reverting to their language
in mid-seminar and nobody dared to point it out.
Nambu was already recognised
as of extraordinary creativity. Because he knew two of my coauthors
were senior Swedish theorists, both at some time members
and chairs of the physics Nobel committee, Peter contacted me in
the 1990s about Nambu being recognised in Stockholm. Eventually Nambu was so rewarded, decades
later then he deserved, in 2008.

\bigskip

\noindent
Peter Freund was extremely respectful of Nambu when both contributed
in our daily tea meetings, with Freund loquacious and
Nambu relatively silent. Yet when Nambu did say something it was 
usually decisive because he was almost always correct.

\bigskip

\noindent
Peter Freund made fundamental contributions about
phenomenological duality and compactifications of
extra spatial dimensions. The two papers I coauthored
with him\cite{PHFPGOF,EFFG} are interesting
and his daily brilliance left a memorable intellectual 
legacy with all who had the good
fortune to meet and discuss with him.

\end{document}